\documentclass[printer]{aa}
\usepackage{multicol}
\usepackage{graphicx}
\usepackage{url}
\usepackage{subcaption}
\usepackage{psfrag}
\usepackage{hyperref}
\DeclareGraphicsExtensions{.pdf,.png,.jpg,.ps,.eps}
\usepackage{natbib}
\usepackage{amsmath}
\usepackage[varg]{txfonts}
\usepackage{booktabs}
\usepackage{amssymb}
\usepackage{wasysym}
\usepackage{longtable}
\bibpunct{(}{)}{;}{a}{}{,} 

\usepackage[normalem]{ulem}
\usepackage{color}

\usepackage{rotating}
\usepackage{longtable}
\usepackage{booktabs}
\usepackage{graphicx} 

\usepackage{graphicx}
\usepackage{txfonts}
\usepackage{lipsum}
\usepackage{subcaption}         
\usepackage{lscape}             
\usepackage{placeins}           
\usepackage[labelsep=period]{caption}
\usepackage{hyperref}
\begin{document}

   \title{The extinction distances for over a thousand planetary nebulae with Gaia measurements}

   
   \author{Juan Deng\inst{1,2,3,4} \and Shu Wang\inst{3}\fnmsep\thanks{Corresponding author: shuwang@nao.cas.cn} \and Biwei Jiang\inst{1,2}\fnmsep\thanks{Corresponding author: bjiang@bnu.edu.cn}\and Licai Deng\inst{3}}

\institute{School of Physics and Astronomy, Beijing Normal University, Beijing 100875,  People's Republic of China
            \and Institute for Frontiers in Astronomy and Astrophysics,
	Beijing Normal University,  Beijing 102206,  People's Republic of China
    \and CAS Key Laboratory of Optical Astronomy, National Astronomical Observatories,
	Chinese Academy of Sciences, Beijing 100101,  People's Republic of China
    \and Department of Astronomy, China West Normal University, Nanchong 637000,  People's Republic of China}

   \date{Received xx, xx}

 
  \abstract
   {Although Gaia has identified the central stars of planetary nebulae (CSPNe) for about 70\% of known Galactic planetary nebulae (PNe), reliable distance estimates remain highly incomplete, with fewer than one quarter having accurate parallaxes. Meanwhile, the classical extinction–distance sample has long been limited to about 70 objects, accounting for only 1.8\% of the Galactic PNe population.} 
   {We aim to obtain a large and homogeneous catalogue of PN distances by refining extinction–distance measurements with Gaia DR3, providing a complementary method to CSPN-parallax-based distances.} 
   {We developed a Gaia-based extinction–distance method for PNe by combining an improved blue-edge approach with an extinction-jump model. Planetary nebula distances were inferred from stellar extinction jumps in line-of-sight extinction–distance profiles and constrained by comparisons with published distances, stellar spatial distributions relative to the PN centre, and the PN radius–distance relation.}
   {We obtain distances for 1,066 PNe, with a median relative uncertainty of 13\% and below 20\% for about 87\% of the sample. This sample includes 765 objects whose CSPN parallaxes have relative uncertainties greater than 20\% and 128 objects without CSPN parallaxes. Our method not only complements CSPN parallax–based approaches for PN distance determination but also extends the traditional extinction-based approach to higher Galactic latitudes. In cases where published distance estimates for the same PN differ significantly, the method helps identify the more reliable distance. In addition, it helps evaluate the reliability of CSPN identifications. We find a likely misidentification in the reported CSPN for Fr2–36, and further analyse 33 PNe with two different CSPNe identifications, suggesting a more suitable CSPN for 15 objects. The resulting catalogue is the largest homogeneous set of extinction-based PN distances to date and provides a robust benchmark for studies of Galactic structure, PN populations, and interstellar extinction.}
  {}


   \maketitle

\section{Introduction} \label{sec:Introduction}

Planetary nebulae (PNe) are short-lived ionized shells ejected by low- and intermediate-mass stars as they evolve from asymptotic giant branch (AGB) to white dwarfs \citep{1957Shklovsky,2013Jacob}. As key tracers of stellar evolution, chemical enrichment, and the interstellar medium \citep{2022Kwitter}, accurate distances to PNe are crucial for determining their intrinsic properties \citep{1986Gathier,1996Pottasch,2022Parker}. However, obtaining reliable PNe distances remains a long-standing challenge, as existing methods struggle to achieve both wide applicability and high measurement reliability.

Distances to PNe are commonly derived through individual or statistical approaches. 
The individual approach measures the distance to a single PN via trigonometric or expansion parallaxes, photoionization modelling, kinematics, cluster membership, or extinction. Although generally regarded as the primary method owing to its higher reliability, it applies to only very few PNe under specific observational or physical conditions \citep{2011Phillips,2022Kwitter,2022Parker}.
The statistical approach infers distances indirectly for large PNe samples from assumed physical properties and empirical relations. While well-suited to population studies, the statistical method introduces systematic uncertainties from calibration datasets, methodologies, and assumptions about PNe evolution and is commonly regarded as a secondary method in the literature \citep{2015Smith,2022Kwitter}.

The high-precision Gaia mission has greatly advanced PNe studies. About 70\% of the $\sim$3,800 known Galactic PNe (\citealp{2016Frew}, hereafter \hypertarget{F16}{F16}) now have identified central stars of planetary nebulae (CSPNe; \citealp{2021Chornay}, hereafter \hypertarget{CW21}{CW21}; \citealp{2021GS}, hereafter \hypertarget{GS21}{GS21}).
Gaia parallaxes for these CSPNe provide the most direct and reliable distance determinations \citep{2022Parker,2022Kwitter}. Nevertheless, more than 75\% of CSPNe are too faint for precise parallaxes, and misidentifications remain non-negligible, limiting the overall utility of CSPNe-based distances. To expand distance coverage, several recent studies have combined CSPN parallaxes with statistical scales, producing samples of 96 PNe (\citealp{2022Ali}, hereafter \hypertarget{A22}{A22}), 843 PNe (\citealp{2023BS}, hereafter \hypertarget{BS23}{BS23}) and 2,211 PNe (\citealp{2024HJMM}, hereafter \hypertarget{HJ24}{HJ24}). The largest of these, \hyperlink{HJ24}{HJ24}, compiled distances from CSPN parallaxes and multiple statistical scales, adopting a rule-based scheme to select the most reliable value for each PN. The associated uncertainties were assigned empirically from comparisons among different distance estimates, ranging from a minimum of 38\% to typical values around 70\% in the published catalogue. Large discrepancies between distances derived from different methods further highlight the difficulty of establishing a reliable distance scale.
For instance, the reported distance to Abell 38 ranges from 474 pc to 6201 pc (\hyperlink{F16}{F16}; \hyperlink{CW21}{CW21}; \hyperlink{GS21}{GS21}; \hyperlink{BS23}{BS23}; \hyperlink{HJ24}{HJ24}), a spread of nearly a factor of 13. Even the most reliable CSPNe-based distances can disagree substantially. For example, the distance to PN HaTr 14 is reported as 12,124 pc and 4,954 pc (\hyperlink{CW21}{CW21}; \hyperlink{GS21}{GS21}), differing by roughly a factor of 3. This illustrates that substantial discrepancies can persist even among results from the most trusted method, most likely due to CSPN misidentification. Thus, even with many identified CSPNe, reliable distances still remain challenging, calling for improved methods.

Among individual methods, the extinction–distance approach is physically well-motivated and independent of assumptions about nebular evolution. Traditionally, PN extinction is inferred from emission-line or radio flux ratios, assuming a fixed total-to-selective extinction optical ratio ($R_V \approx 2.5$–3.2; e.g. \citealt{1978Acker,1995Saurer}), with typical values of \(A_V \sim 0.1\text{–}8.5~\mathrm{mag}\) (\citealp{2021Dharmawardena}, hereafter \hypertarget{D21}{D21}). Distances are then estimated either by comparison with field stars along the same sightline \citep{1986Gathier} or by matching to 3D extinction maps (\hyperlink{D21}{D21}). Despite decades of effort, this method has produced distances for only a few dozen PNe (6–70 objects; \citealt{1973Lutz,1978Acker,1994Martin,1995Saurer,2011Giammanco}, hereafter \hypertarget{G11}{G11}; \hyperlink{D21}{D21}). The main limitations of this method arise from incomplete or uncertain PN extinction estimates, which result from restricted Balmer coverage, missing or low-quality radio fluxes, uncorrected intrinsic colours, and the use of a constant $R_V$ that in reality varies significantly across sightlines (\hyperlink{D21}{D21}, \citealp{2014Wang,2019wangshu,2023wang,2025Green}). Its applicability is further constrained by the small number of comparison stars and by the coarse angular resolution of current extinction maps. 

Recent advances offer a more robust foundation for extinction-based distances. The blue-edge method, used to estimate stellar intrinsic colours, has been continuously refined \citep{2014Wang,2019wangshu,2024zhoahe,2025Deng}, yielding accurate stellar extinction without assuming a fixed $R_V$. The extinction-jump model enables distance estimates via extinction jumps and has been successfully applied to supernova remnants and molecular clouds (\citealt{2020wangshu,2024caozhetai,2025chenxiaohan}). 
When combined with Gaia’s all-sky coverage and high-precision parallaxes, which supply large and reliable stellar samples, these developments enable substantially improved extinction–distance determinations for PNe. 

In this work, we combined the refined blue-edge method with the extinction-jump model to develop a Gaia-based extinction–distance method for PNe with identified CSPNe. The resulting distances are constrained jointly by multiple indicators, including CSPN-based estimates, parallaxes, and statistical scales, and are further validated using 3D extinction maps and the empirical $R_{\rm PN}$–distance relation. This method yields robust distances for over one thousand PNe, addresses the difficulties posed by missing or unreliable CSPNe parallaxes, and helps identify more reliable distances when published estimates differ widely.
This paper is organized as follows.  
Section~\ref{sec:data_samples} describes the sample and data. Section~\ref{sec:Method} presents the methodology. Section~\ref{Step4} discusses validation. Section~\ref{sec:Results and Discussion} provides the results and discussion, and Section~\ref{Conclusions} summarises our conclusions.

\section{Sample and data}
\label{sec:data_samples}
\subsection{PN sample}
\label{samples}

\begin{figure*}[h!]
\centering
\includegraphics[width=0.9\hsize]{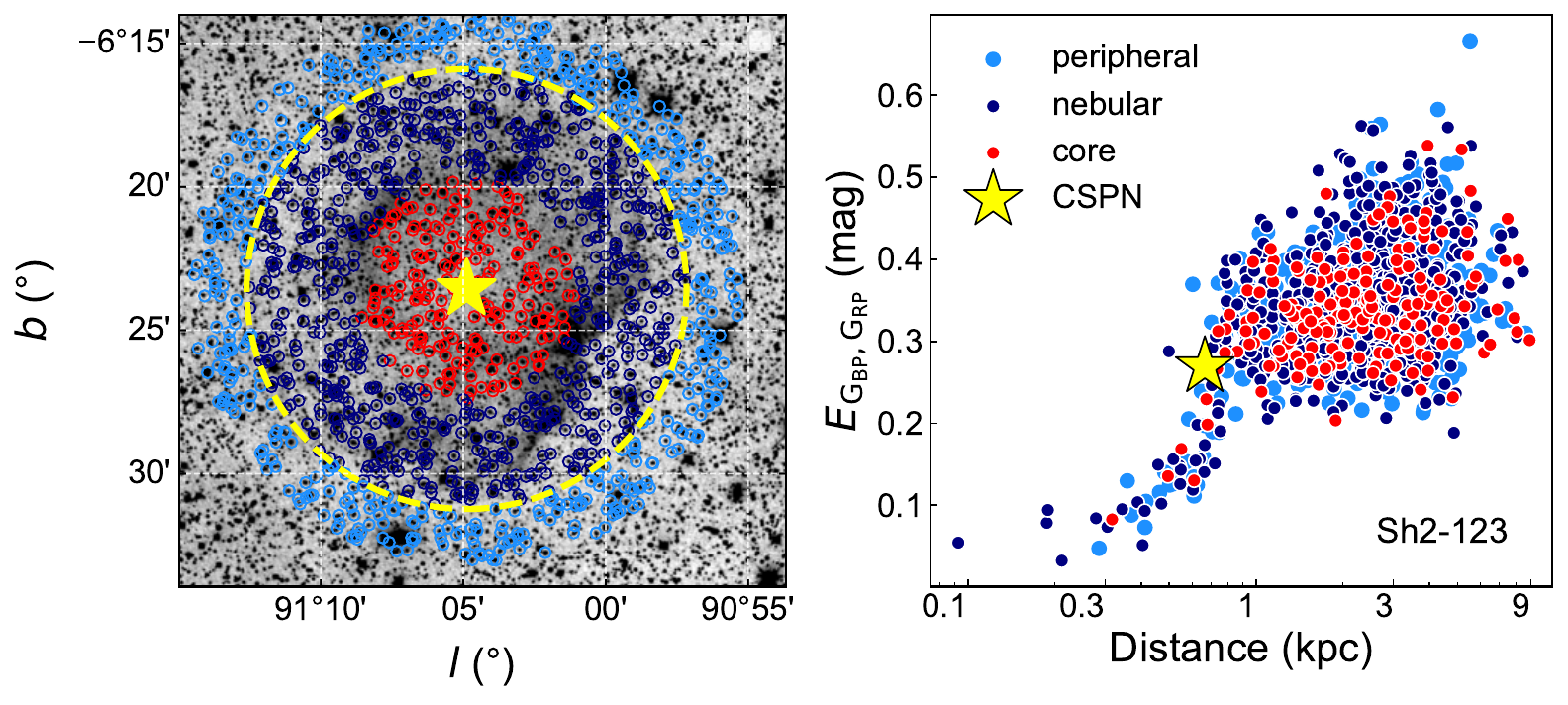}
\caption{Left: Stars overlaid on the Digitized Sky Survey image with $R_\mathrm{PN}$ (dashed yellow circle), the CSPN (yellow star), as well as core, nebular, and peripheral sources (red, dark blue, and blue dots). Right: Extinction vs distance.}
\label{PN_spatial_slice}
\end{figure*}

To address the shortage of reliable distances for PNe with identified CSPNe, we constructed our working sample from the two largest available CSPNe catalogues, \hyperlink{CW21}{CW21} and \hyperlink{GS21}{GS21}. Both catalogues are based on the HASH (Hong Kong/AAO/Strasbourg/H$\alpha$) PN database \citep{2016HASH}, which includes nearly all known Galactic PNe from previous compilations and provides CSPNe identifications using Gaia EDR3. \hyperlink{CW21}{CW21} compiled a catalogue of 2,117 CSPNe and derived distances for 733 PNe by combining Gaia parallaxes with the widely used statistical distance catalogue of \hyperlink{F16}{F16}. \hyperlink{GS21}{GS21} reported 2,035 CSPNe, among which 405 distances from \citeauthor{2021Bailer-Jones} (2021, hereafter \hypertarget{BJ21}{BJ21}) were classified as reliable and recommended for use.
Combining both catalogues yields 2,570 Galactic PNe including all CSPNe, with 897 PNe having reliable distances. 

For these 2,570 PNe, we retained all reported distances and extracted their central coordinates and angular radii ($R_\mathrm{PN}$) from the HASH database. Each PN was assigned an individually defined circular field for extracting stars to construct a reliable extinction–distance profile (see Section \ref{search regions}). Planetary nebulae lacking a defined $R_\mathrm{PN}$ or containing fewer than ten stars within the field were excluded, leading to the final sample of 2,225 PNe.

\subsection{Data}\label{data}
For each PN, we selected stellar samples along the line of sight to trace the cumulative extinction towards it.
Gaia DR3 provides high-precision photometry in the $\mathit{G_\mathrm{BP}}$, $\mathit{G_\mathrm{RP}}$, and $\mathit{G}$ bands, together with spectroscopic and astrometric data for over 1.8 billion sources \citep{2023Gaia3}.
From its low-resolution XP spectra, \citet{2023Andrae} derived robust stellar atmospheric parameters for $\sim$175 million stars, including effective temperature ($T_{\rm eff}$), surface gravity ($\log g$), and metallicity ([M/H]), with typical uncertainties of $\sim$50 K, 0.08 dex, and 0.1 dex, respectively.
For this study, we extracted these stellar parameters from \citet{2023Andrae}, as they are required for intrinsic-colour estimation using the blue-edge method. Stellar distances were taken from \hyperlink{BJ21}{BJ21}, which provides geometric distances from Gaia parallaxes and photogeometric distances incorporating photometry and astrophysical priors, yielding improved accuracy for stars with low-quality parallaxes. Distances for CSPNe were adopted from either the geometric or photogeometric solution depending on parallax quality, whereas non-CSPNe used only geometric distances. We excluded stars that satisfied any of the following criteria: 
\begin{itemize}
    \item negative or missing parallaxes,
    \item fractional parallax error $>$20\%,
    \item $G>17.65$ mag,
    \item $T_{\rm eff}$ outside the range $4000$–$8000$ K,
    \item uncertainty in $\log g$ or [M/H] $>$ 0.3,
    \item $T_{\rm eff}$ uncertainty $>$ 0.1 (\hyperlink{BJ21}{BJ21}, \citealp{2023Andrae,2024zhoahe}).
\end{itemize}
Combined with the depth and precision of Gaia DR3, these selection criteria ensured a clean and reliable stellar sample, providing a solid foundation for constructing extinction–distance profiles.

\subsection{PN stellar field}\label{search regions}
The extinction–distance method requires a stellar sample for each PN to build an extinction–distance profile along its line of sight. Typically, a stellar field is defined around each PN to select nearby stars. The field size must balance statistical robustness with sensitivity to extinction jumps. 
Previous studies typically adopted constant stellar fields on angular scales ranging ($R_\mathrm{field}$) from $1.5^\circ$ to $10'$ (\citealp{1973Lutz}; \hyperlink{G11}{G11}) to ensure sufficiently large stellar samples. Although simple, this strategy ignores PN size variations (typically $\sim$1$'$–2.5$'$) and introduces background contaminants, which reduce spatial sensitivity and limit the applicability of the method.

Using Gaia DR3, we defined $R_\mathrm{field}$ for each PN with a PN-size–dependent approach guided by spatial-consistency checks to construct reliable stellar samples and minimise contamination from nearby regions. The planetary nebula angular radii ($R_\mathrm{PN}$) were taken from the HASH database \citep{2016HASH}. Stars within each $R_\mathrm{field}$ were then divided into three concentric zones based on spatial distribution. The core zone includes stars within 0.5 $R_\mathrm{PN}$, the nebular zone covers 0.5–1 $R_\mathrm{PN}$, and the peripheral zone extends from $R_\mathrm{PN}$ to $R_\mathrm{field}$. This division enabled us to check the spatial consistency of extinction–distance trends and validating the chosen $R_\mathrm{field}$. 

As shown for Sh2-123 (\autoref{PN_spatial_slice}), the extinction–distance relations across these zones are consistent, confirming the reliability of our $R_\mathrm{field}$ selection.
Systematic tests across the full sample yield the optimised selection criteria summarised in \autoref{R_search}. For compact PNe, $R_\mathrm{field}$ was extended to several times $R_\mathrm{PN}$. These criteria were designed to maximise completeness while minimising contamination.

\begin{table}[h]
\centering
\caption{PN-size–dependent field radius ($R_{\rm field}$) scaling.}
\label{R_search}
\begin{tabular}{lc}
 \hline \hline
$R_{\mathrm{PN}}$ (arcsec) & $R_{\mathrm{field}}$ (arcsec) \\
\hline
$<5$        & $3.5 \times R_{\mathrm{PN}}$  \\
$5$–$100$   & $2.5 \times R_{\mathrm{PN}}$  \\
$100$–$300$ & $1.5 \times R_{\mathrm{PN}}$   \\
$>300$      & $R_{\mathrm{PN}} + 100$ \\
\hline
\end{tabular}
\tablefoot{The $R_{\rm field}$ is scaled from the PN angular radius ($R_{\rm PN}$) to balance background star coverage and contamination avoidance.}
\end{table}

\section{Method} \label{sec:Method}
We adopted a three-step approach to estimate PN distances based on extinction–distance profiles. Extinction–distance profiles were first constructed for each PN sightline. Significant extinction jumps were then identified in these profiles through model fitting or by directly tracing features near the CSPNe. Finally, distances inferred from extinction jumps were adopted as PN distances only after satisfying multiple independent consistency checks.

\subsection{ Extinction–distance profile construction}\label{step1}

The first and most critical step is to determine the extinction for each star within $R_{\rm field}$, expressed as
\begin{equation}
E_\mathrm{G_{BP}, G_{RP}} = (G_\mathrm{BP} - G_\mathrm{RP}) - (G_\mathrm{BP} - G_\mathrm{RP})_0,
\end{equation}
where the observed colour $(G_\mathrm{BP} - G_\mathrm{RP})$ is taken from Gaia DR3 photometry, and the intrinsic colour $(G_\mathrm{BP} - G_\mathrm{RP})_0$ is obtained via the widely used blue-edge method \citep{2014Wang,2019wangshu,2024Wangshu,2025Deng}, which estimates intrinsic colours from stellar atmospheric parameters ($T_{\rm eff}$, $\log g$, and [M/H]). 
We adopted the machine-learning-enhanced blue-edge method of \citet{2024zhoahe}, with a typical intrinsic-colour uncertainty of 0.014 mag. The derived extinction values do not rely on a fixed $R_V$ and offer improved accuracy over previous methods. Stellar distances were taken from \hyperlink{BJ21}{BJ21}. These extinction and distance estimates form the basis of the extinction–distance profiles used in the subsequent analysis.

\begin{figure*}[h!]
\centering
\includegraphics[width=0.9\hsize]{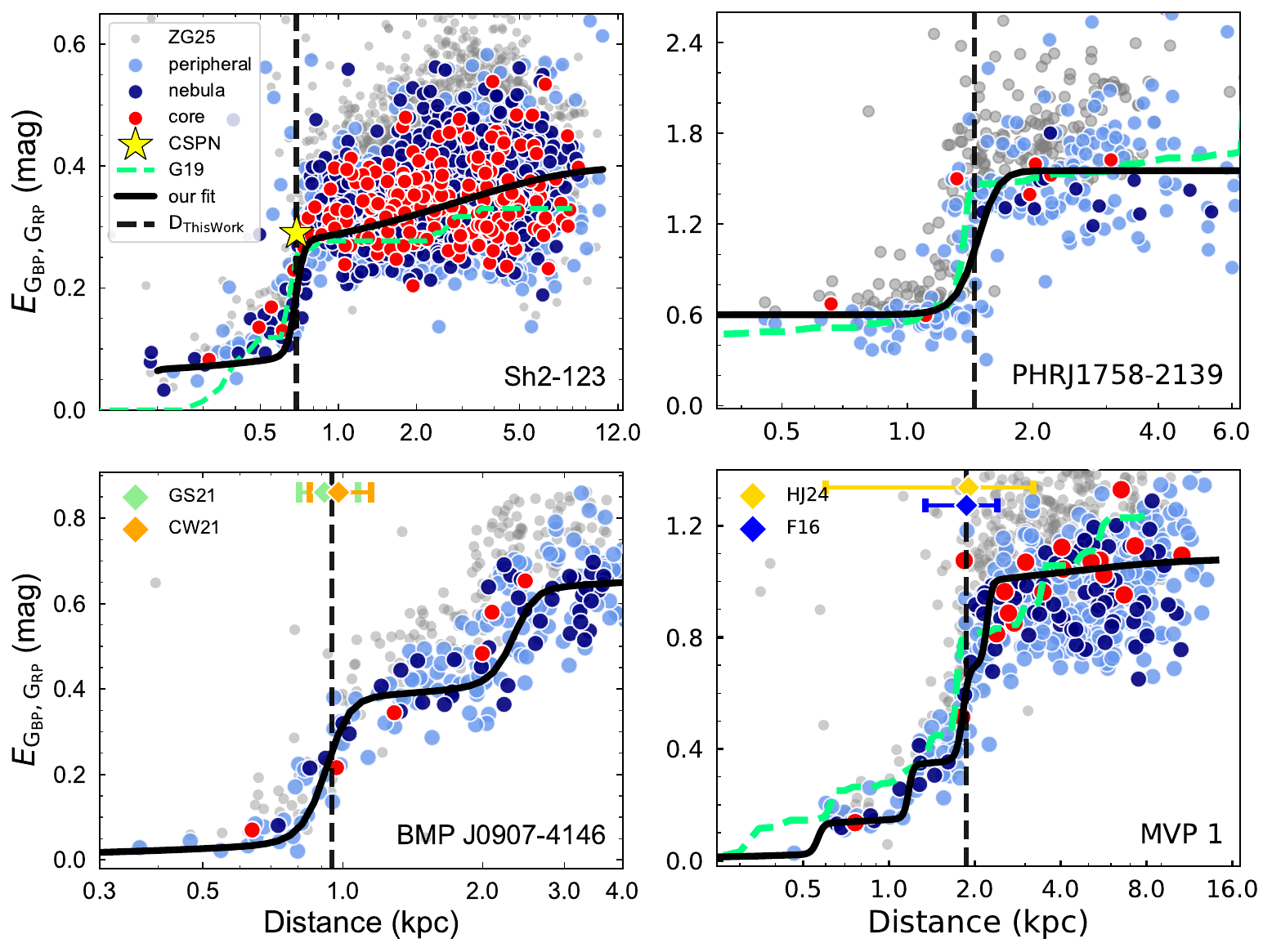}
\caption{Extinction–distance profiles for four PNe with single jumps in the top row and multiple jumps in the bottom row. PN names are labelled at the lower right of each panel. Distance (in kiloparsec) vs extinction $E_\mathrm{G_{BP}, G_{RP}}$ (in magnitude). The source colours are the same as those in Figure 1, and the solid black lines show the fits obtained with our extinction-jump model. PN positions are shown with vertical dashed black lines. The grey points represent ZG25 stars, and the dashed green lines indicate G19 curves. PN distance constraints are shown as diamonds with error bars from GS21 (light green), CW21 (orange), HJ24 (yellow), and F16 (blue). Conventions are consistent throughout.}
\label{combined_2x2}
\end{figure*}

\subsection{Extinction jump identification}\label{Step2}
We identified significant extinction jumps in the extinction–distance profiles using an extinction-jump model. These jumps appear as sharp increases in extinction over narrow distance ranges and define extinction-jump distances. Planetary nebulae can produce such detectable extinction features through dust within the nebula itself and in its surrounding extended AGB dust, whose spatial extent far exceeds that of the ionised nebula.

Our model, following \citet{2017chenBQ}, \citet{2020wangshu}, \citet{2020zhaohe}, \citet{2021sunmingxu}, and \citet{2025chenxiaohan}, assumes that cumulative extinction along the line of sight arises from two components: diffuse interstellar dust and PN-associated dust.
The total extinction at distance $\mathit{d}$ is then expressed as
\begin{equation}
A(d) = A_0(d) + A_1(d),
\end{equation}
where \(A_0(d)\) represents the foreground diffuse extinction and
\(A_1(d)\) accounts for the PN-associated extinction. 
For PNe, we approximated the foreground extinction as a constant, \(A_0(d) = A_0\), following the treatment adopted for nearby molecular clouds by \citet{2023cao}. This approximation is motivated by the weak foreground extinction towards nearby PNe and the limited number of foreground stars, which makes a distance-dependent foreground difficult to constrain. 
The parameter \(A_0\) was fitted as a free parameter and represents the foreground extinction level prior to the jump.
The PN-associated extinction is modelled as
\begin{equation}
A_1(d) = 
\Delta A \, \frac{1 + \mathrm{erf} \!\left[ \frac{d - d_0}{\delta d} \right]}{2},
\end{equation}
where \(\Delta A\) is the amplitude of the extinction jump and \(d_0\) the distance to the PN centre. Here, \(\delta d\) is set to the PN diameter as a smoothing scale, but in practice it mainly reflects the propagation of stellar distance uncertainties (see Section \ref{subsec:PN distance}). The fitting parameters are \(A_0\), \(\Delta A\), and \(d_0\).
We employed Markov chain Monte Carlo (MCMC) to fit the extinction–distance profiles, deriving best-fit parameters \((A_0,\Delta A, d_0)\).
Parallax uncertainties were propagated into the distance errors and incorporated into the MCMC likelihood. The uncertainty associated with each candidate distance solution was quantified by the $1\sigma$ statistical uncertainty of the fitted jump location $d_0$, as inferred from the MCMC posterior distribution. An extinction jump was considered significant when $\Delta A$ > 0.07 mag (five times the intrinsic colour uncertainty) and the distance parameter $d_0$ was well constrained with a single dominant peak in the posterior distribution.
For cases with multiple peaks in the MCMC posterior indicating several extinction jumps, multiple components were fitted simultaneously.

\subsection{PN distance estimation}
\label{subsec:Step3}
In the previous step, each significant extinction jump identified in the extinction–distance profile yields an extinction–jump distance. Not all extinction jumps detected along a given line of sight are associated with the PN, as similar extinction features may also arise from interstellar clouds. Thus, these distances were evaluated using available comparisons with 3D dust extinction maps, analysis of the stellar spatial distribution relative to the PN centre, and the use of published distance constraints with different priority levels. The final adopted PN distance was selected based on the combined results of these checks.

All profiles were compared with two representative 3D interstellar dust extinction maps. The first is the recent map from \citeauthor{2025zhangxiangyu} (2025; hereafter \hypertarget{ZG25}{ZG25}), which provides extinction curves, individual stellar extinction values, and distances for 130 million stars. The second is the widely used map from \citeauthor{2019Green} (2019; hereafter \hypertarget{G19}{G19}), covering three-quarters of the sky and providing extinction–distance curves at $3.4^{\prime}$–$13.7^{\prime}$ resolution.
Within each PN’s $R_\mathrm{field}$, \hyperlink{ZG25}{ZG25} single-star data were compared point-by-point with our measurements, while \hyperlink{G19}{G19} curves were used to verify overall trends and jump locations at larger scales.

For profiles exhibiting a single extinction jump, the distance was considered reliable when the jump was further supported either by agreement with the CSPN distance or by a strong extinction increase traced by stars nearest the PN centre.
As illustrated in \autoref{combined_2x2}, when a CSPN distance is available (top left panel), our extinction–distance measurements (coloured dots and black line) are consistent with the discrete \hyperlink{ZG25}{ZG25} data (grey dots) and the \hyperlink{G19}{G19} curve (green line) and are further supported by the CSPN distance.  When no CSPN distance is available (top right panel), the extinction jump is instead supported by a clear extinction increase traced by stars closest to the PN centre (red dots, $\Delta E_\mathrm{G_{BP}, G_{RP}} = 1.0$ mag), providing an independent indication of the PN location.

When multiple extinction jumps produce more than one extinction-jump distance, these distances were further evaluated using distance constraints with different priority levels. These constraints are based on multiple independent distance estimates, together with their associated uncertainties. The highest-priority indicators are CSPN-based distances (\hyperlink{CW21}{CW21}, \hyperlink{GS21}{GS21}). When available, no other constraints were considered. The next priority is parallax-based CSPN distances (\hyperlink{BJ21}{BJ21}). The remaining indicators include canonical statistical distances (\hyperlink{F16}{F16}, with mean values adopted following \hyperlink{CW21}{CW21}), CSPN-calibrated statistical distances (\hyperlink{A22}{A22}, \hyperlink{BS23}{BS23}), and the latest distances (\hyperlink{HJ24}{HJ24}), all considered of equal weight. When indicators carry equal weight, distances supported by more independent constraints are considered more reliable.
The bottom panels of \autoref{combined_2x2} present two examples. The lower-left panel shows a PN exhibiting two extinction jumps along the line of sight, at $\sim$1 kpc ($\Delta E_\mathrm{G_{BP}, G_{RP}}=0.4$ mag) and $\sim$2.5 kpc ($\Delta E_\mathrm{G_{BP}, G_{RP}}=0.2$ mag) respectively. The first jump is identified as the PN location based on the highest-priority indicators (\hyperlink{CW21}{CW21}, \hyperlink{GS21}{GS21}). The lower-right panel shows the high-confidence region where the \hyperlink{HJ24}{HJ24} and \hyperlink{F16}{F16} constraints overlap, with the core-source jump clearly detected ($\Delta E_\mathrm{G_{BP}, G_{RP}}=1.0$ mag), confirming the PN location.

Throughout this paper, our results are shown in black and literature results in colour, with stellar distances as dots and cloud distances as diamonds. The colour coding assigns yellow to \hyperlink{HJ24}{HJ24}, cyan to \hyperlink{BS23}{BS23}, purple to \hyperlink{A22}{A22}, orange to \hyperlink{CW21}{CW21}, light green to \hyperlink{GS21}{GS21}, blue to \hyperlink{F16}{F16}, and grey to \hyperlink{BJ21}{BJ21}.
In rare cases where the distance constraints are insufficient to estimate the PN distance, the $R_{\mathrm{PN}}$–distance relation offers auxiliary validation (Section \ref{subsec:scale}).

\section{Method validation}\label{Step4}
\begin{figure}[h!]
\centering
\includegraphics[width=0.9\hsize]{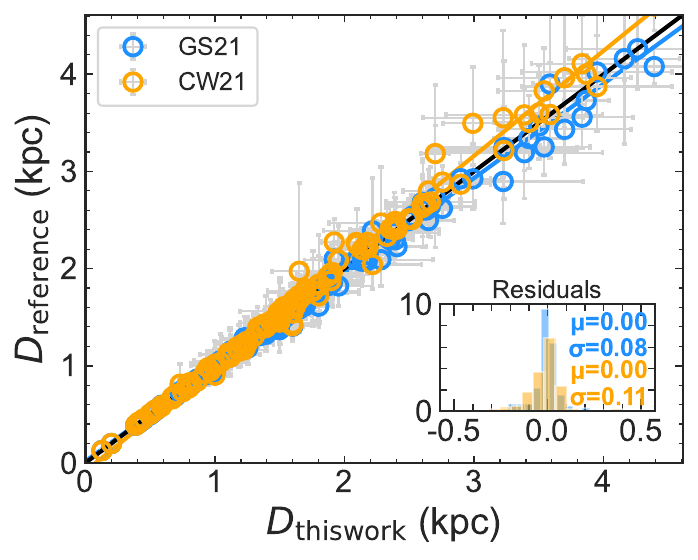}
\caption{Comparison of distances for the tight sample. Distances derived in this work are compared with those from \protect\hyperlink{CW21}{CW21} (orange circles) and \protect\hyperlink{GS21}{GS21} (blue circles). All data points are shown with their associated uncertainties. The solid black line shows $y=x$, while the coloured lines represent the best-fitting relations. The inset displays the distribution of residuals.}
\label{fig:high_confidence}
\end{figure}

To evaluate the reliability of our method, we selected a tight sample of 137 PNe with both \hyperlink{CW21}{CW21} and \hyperlink{GS21}{GS21} distances consistent within 3$\sigma$. Distances derived for this sample using our method were then compared with the reference values, yielding residuals with a mean of 0.00 and dispersions of approximately 0.08 kpc for GS21 and 0.11 kpc for CW21, confirming the robustness of our distance estimates (\autoref{fig:high_confidence}).

\begin{table}[h]
\centering
\caption{Individual distance estimates for PNe}
\label{othermethod1}
\begin{tabular}{lccc}
\hline \hline
Name & $D_{\mathrm{thiswork}}$ & $D_{\mathrm{Ref}}$ & Method \\ & (kpc)&(kpc)&\\
\hline
We~2-5     & $2.1 \pm 0.5$   & $2.3 \pm 0.6$   & Kinematic method$^{1}$ \\
CVMP~1     & $1.4 \pm 0.4$   & $1.9 \pm 0.7$   & Kinematic method$^{2}$ \\
Hen~2-11   & $1.6 \pm 0.8$   & $1.9 \pm 0.6$   & Kinematic method$^{3}$ \\
HFG~2      & $1.9 \pm 0.2$   & $1.9 \pm 0.6$   & Kinematic method$^{4}$ \\
NGC~5189   & $1.5 \pm 0.1$   & $1.0 \pm 0.7$   & Kinematic method$^{5}$ \\
SuWt~2     & $2.1 \pm 0.1$   & $2.3 \pm 0.6$   & Kinematic method$^{6}$ \\
Abell~79   & $3.9 \pm 0.6$   & $4.4 \pm 1.0$   & Kinematic method$^{7}$ \\
DPV~1      & $4.9 \pm 0.5$   & $3.8 \pm 1.1$   & Outburst brightness$^{8}$ \\
Hen~1-5    & $3.0 \pm 0.1$   & $2.8 \pm 0.8$   & Outburst brightness$^{8}$ \\
FG~Sge     & $3.0 \pm 0.1$   & $2.5 \pm 0.5$   & Pulsation theory$^{9}$ \\
NGC~7293   & $0.2 \pm 0.02$ & $0.18 \pm 0.03$ & Proper motion$^{10}$ \\
Sh~2-188   & $0.9 \pm 0.05$  & $0.85 \pm 0.5$  & Convergent parallax$^{11}$ \\
Abell~63   & $2.6 \pm 0.2$   & $2.4 \pm 0.4$   & Eclipsing binary$^{12}$ \\
SuWt~2     & $2.1 \pm 0.06$  & $2.3 \pm 0.2$   & Eclipsing binary$^{13}$ \\
HFG~1      & $0.69 \pm 0.08$ & $0.63 \pm 0.32$ & Eclipsing binary$^{14}$ \\
\hline
\end{tabular}
\tablefoot{
$D_{\mathrm{Ref}}$ denotes distances adopted from the literature.
Superscripts indicate the reference sources:
$^{1}$\cite{2012Kinematic},
$^{2}$\cite{1997Corradi},
$^{3}$\cite{1989Meaburn},
$^{4}$\cite{2008Frew},
$^{5}$\cite{2012NGC5189},
$^{6}$\cite{2010SuWt},
$^{7}$\cite{2001R},
$^{8}$\cite{2016Frew},
$^{9}$\cite{1980MA},
$^{10}$\cite{1984Eggen},
$^{11}$\cite{2006Wareing},
$^{12}$\cite{1994Abell63},
$^{13}$\cite{2010E10},
$^{14}$\cite{2005E05}.
}
\end{table}

\begin{figure}[h!]
\centering
\includegraphics[width= 0.8\hsize]{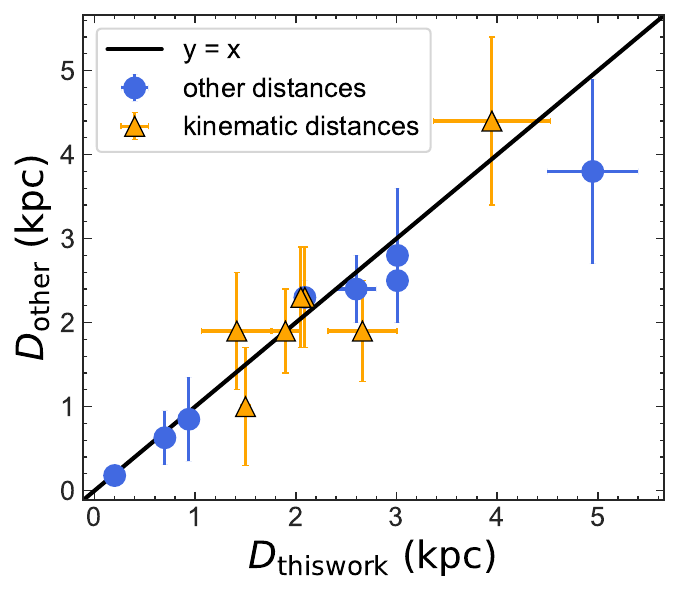}
\caption{Comparison of our distances with selected individual PN measurements from \autoref{othermethod1}. The yellow triangles indicate kinematic distances. The blue circles denote other independent methods.}
\label{othermethod}
\end{figure}

In addition to tight-sample validation using CSPNe distances, we also include other representative individual PN distance estimates, such as kinematic, outburst-brightness, pulsation-based, and convergent parallax distances.
Kinematic distances, derived from sky position and radial velocity under an assumed Galactic rotation model, are particularly suited for young, disk PNe (typically Type I). As in \hyperlink{F16}{F16}, we adopted the best available kinematic distances (e.g. \citealp{2010SuWt,2012NGC5189}). Several historically observed final-flash CSPNe were measured via the outburst-brightness method (\hyperlink{F16}{F16}). Additional representative distances include FG Sge from pulsation theory \citep{1980MA}, Sh 2-188 from PN–ISM interaction modelling combined with CSPN proper motion \citep{2006Wareing}, and NGC 7293 from convergent parallax assuming Hyades membership \citep{1984Eggen}. We also included three PNe hosting rare eclipsing-binary CSPNe, whose distances were derived from the eclipsing-binary method  \citep{1994Abell63,2005E05,2010E10}.
\autoref{othermethod1} summarises these distances with their methods and references. The strong agreement among these diverse techniques in \autoref{othermethod} further supports the reliability and robustness of our method.

\section{Results and discussion} \label{sec:Results and Discussion}
\subsection{PNe distances} \label{subsec:PN distance}

\begin{figure}
\centering
\includegraphics[width=0.9\hsize]{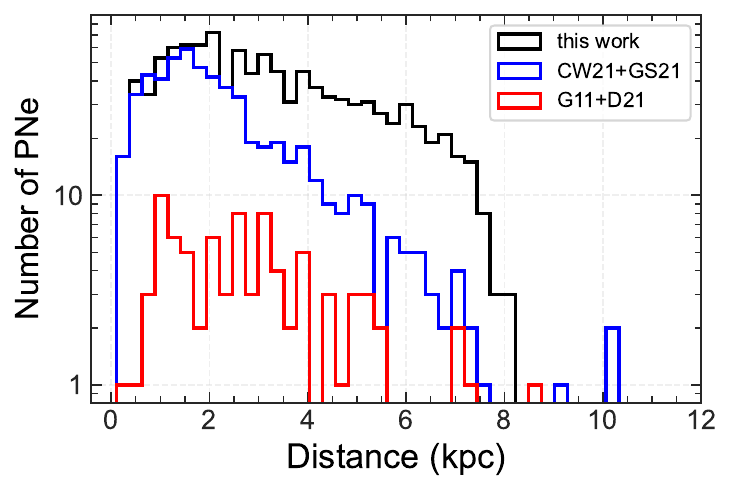}
\caption{Step histograms of PN distance distributions for this work (black), CSPN-based samples (blue), and extinction–distance samples (red).}
\label{hist}
\end{figure}

We applied our method to 2,225 PNe (Section~\ref{samples}), obtaining extinction distances for 1,066 objects, 87\% of which have relative distance uncertainties below 20\%, with a median value of 13\%, as listed in \autoref{alldistances}. To quantify the degree of independent external support, we assigned a quality flag Q (1–3) to each adopted distance: Q = 3 denotes cross-validation by CSPN-based distance estimates together with at least one additional independent distance indicator; Q = 2 denotes support from CSPN-based distance estimates alone or, in the absence of CSPN information, by at least two mutually consistent independent distance indicators; Q = 1 denotes support from only one independent indicator. The Q flag characterizes the strength of cross-validation rather than the intrinsic robustness of the extinction-based distances themselves. Among these, 765 PNe have CSPN parallaxes with relative uncertainties $\geq$20\%, 128 lack CSPN parallaxes, and 124 distances were determined for the first time.
\autoref{hist} compares our distance distribution with CSPN-based distances (\hyperlink{CW21}{CW21} + \hyperlink{GS21}{GS21}, 897 PNe), as well as with the largest-to-date and the most recent extinction–distance samples (\hyperlink{G11}{G11}, 70 PNe and \hyperlink{D21}{D21}, 17 PNe). Although our observational depth remains within the limits of previous work, our results substantially increase the number of distances to PNe derived using individual methods.

\begin{table}[h]
\centering
\caption{Distance estimates for PNe}
\label{alldistances}
\begin{tabular}{lcccc}
 \hline \hline
PNG & Name & $D$ &$D_\mathrm{err}$ & Q \\
 &  & (pc) & (pc) &  \\
\hline
000.6-01.0 & JaSt 77 & 2988 & 587 & 1 \\
001.0+01.4 & JaSt 2-4 & 1342 & 120 & 3 \\
001.0+01.9 & K 1-4 & 2650 & 203 & 2 \\
$\cdots$ & $\cdots$ & $\cdots$ & $\cdots$ & $\cdots$ \\
\hline
\end{tabular}
\tablefoot{
$D$ and $D_\mathrm{err}$ denote the distances derived in this work and their corresponding $1\sigma$ uncertainties. Q indicates the adopted distance quality flag. The full table is available at the CDS via the VizieR online service and on Zenodo.
}
\end{table}

Extinction jumps detected by our method span 
$0.15~\mathrm{mag} < \Delta E_{\mathrm{G_{BP}, G_{RP}}} < 2.8~\mathrm{mag}$, 
consistent with the largest recent PN extinction sample from \protect\hyperlink{D21}{D21} ($\Delta E_{\mathrm{G_{BP}, G_{RP}}} \approx 0.04\text{--}3.6~\mathrm{mag}$ using the 
$A_V$--$\Delta E_{\mathrm{G_{BP}, G_{RP}}}$ relation of \citealp{2019wangshu}).
Planetary nebulae with higher extinction (\(\Delta E_\mathrm{G_{BP}, G_{RP}} >2.8~\mathrm{mag}\)) were not detected by our method, likely due to the paucity of visible stars in such dense regions, marking the method’s detection limits. 

\begin{figure*}[h!]
\centering
\includegraphics[width=\hsize]{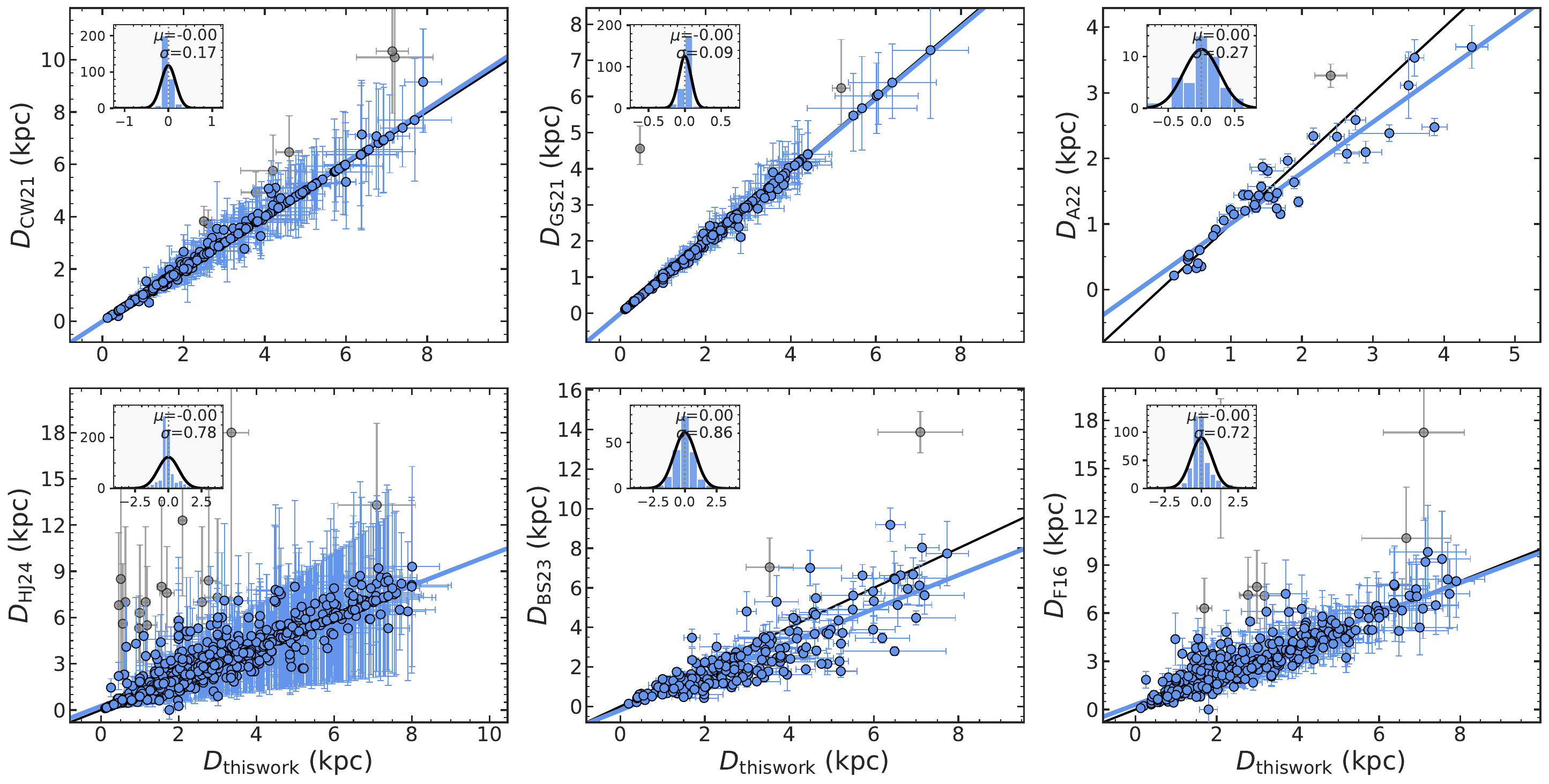}
\caption{Comparison of our distances (x-axis, kiloparsec) with literature values (y-axis, kiloparsec).
Each panel shows a reference scale (\protect\hyperlink{CW21}{CW21}, \protect\hyperlink{GS21}{GS21}, \protect\hyperlink{F16}{F16}, \protect\hyperlink{A22}{A22}, \protect\hyperlink{BS23}{BS23}, and \protect\hyperlink{HJ24}{HJ24}).
The solid black line shows $y = x$, and the coloured lines represent orthogonal linear fits obtained with iterative $3\sigma$ rejection for each dataset. The grey points indicate the objects excluded ($>3\sigma$) from the final fit. The insets show the residual distributions with the mean ($\mu$) and scatter ($\sigma$).}
\label{compare_dis}
\end{figure*}

Planetary nebulae typically have physical radii below 1.5 pc, reaching 2.5–3 pc at late evolutionary stages \citep{2010Frew,2022Parker}, and our results are consistent with this, with a median radius of 0.45 pc and 90\% of the sample spanning 0.12–1.51 pc. The apparent width of extinction jumps in distance space, however, is generally larger than the physical scale of PNe, with a median of about 150 pc, indicating that it reflects the effective distance resolution of the extinction–distance method rather than any intrinsic physical thickness. We performed a simple simulation by introducing a 10\% relative uncertainty in tracer-star extinctions and distance uncertainties of 0\%, 10\%, and 20\% for extinction jumps $\mathit{d}$ at 1–3 kpc. The inferred jump width, $\Delta d$, increases systematically with distance uncertainty, growing from $\sim$6 pc at $d = 1$ kpc with no distance uncertainty to $\sim$110–150 pc for 10\%–20\% uncertainties and reaching $\sim$0.5 kpc at $d = 3$  kpc for a 20\% uncertainty. 
This trend confirms that the jump width primarily reflects the propagation of distance uncertainties and is consistent with our earlier conclusions.

Our distances are compared with literature values in \autoref{compare_dis}. Orthogonal fits were performed with a single 3$\sigma$ rejection and refitting, and the discarded sources are shown in grey. They agree best with CSPN-based results from \hyperlink{GS21}{GS21} and \hyperlink{CW21}{CW21}. \hyperlink{A22}{A22} and \hyperlink{BS23}{BS23} tend to underestimate distances, whereas \hyperlink{HJ24}{HJ24} exhibits a pronounced systematic error growth with distance. \hyperlink{F16}{F16} matches well. The grey outliers in \hyperlink{GS21}{GS21} and \hyperlink{CW21}{CW21} mainly reflect inconsistent CSPN identifications (Section \ref{subsubsec:r-d-3}), while other outliers trace method-related uncertainties (Section \ref{subsubsec:r-d-2}). In general, the final fits are close to one-to-one ($\mu = 0$), with rejected outliers accounting for only 1.5\% of our sample. Our method yields smaller uncertainties while remaining consistent with the literature, thereby providing more robust distance estimates.

To further assess the reliability of our distance estimates for distant PNe, particularly along low-latitude sightlines where extinction is strong, we additionally compared our results with the 3D extinction map of \citet{Marshall2006}. Constructed from 2MASS near-infrared photometry combined with a Galactic stellar population model, this map remains effective in high-extinction regions and therefore provides a targeted complementary test for distant PNe. For a meaningful comparison, a PN must lie within both the sky coverage of the map, and its sampled distance must range along the relevant line of sight. Within this common range, 37 distant PNe with d > 4.5 kpc can be directly compared. Among these, ten coincide with or lie close to clear extinction jumps in the Marshall profiles, while the others fall in flat segments of the curves without a distinct jump. Within this subset, the jump-associated objects tend to have larger angular radii (mostly $> 28''$), whereas the remaining ones are smaller. Given the angular resolution of the map ($15' \times 15'$) and its discrete distance sampling, small-scale extinction structures are expected to be smoothed by spatial averaging. No systematic offset relative to our distance scale is found for this subset.

\begin{figure}[h!]
\centering
\includegraphics[width=1\hsize]{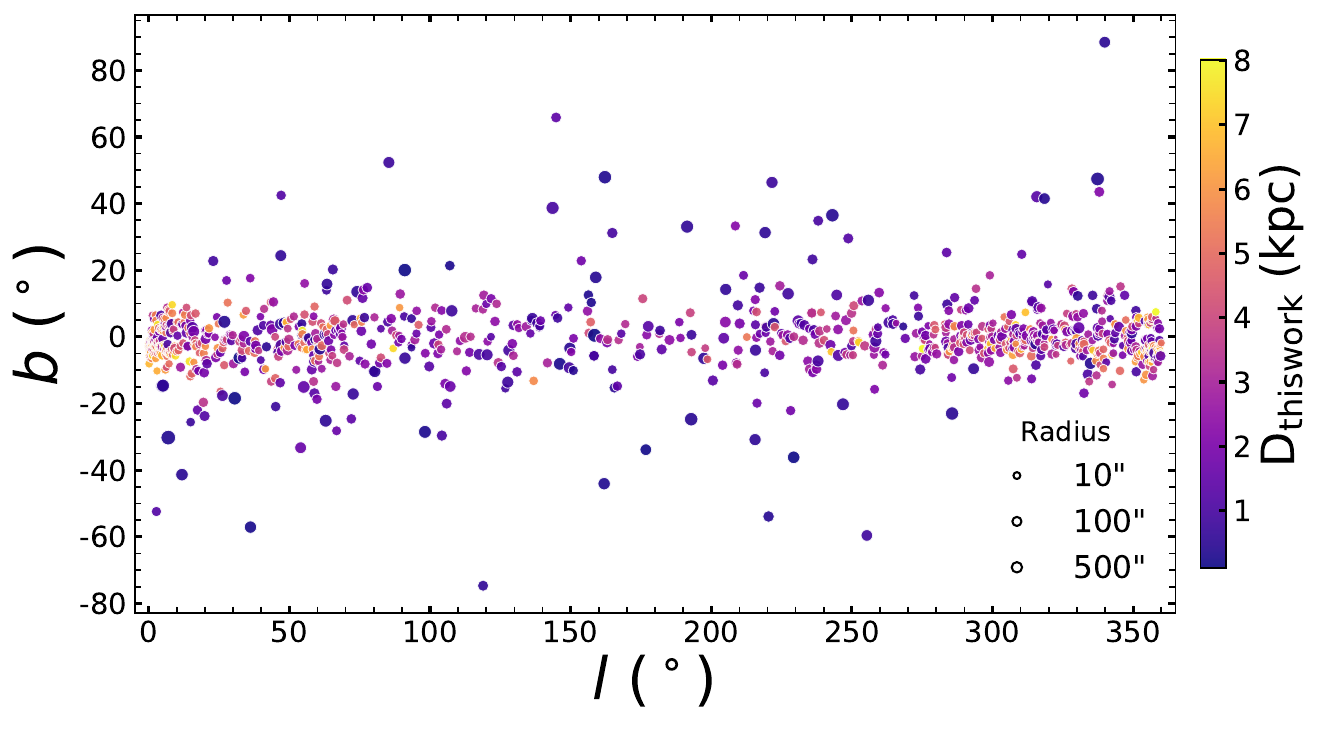}
\caption{Spatial distribution of the PN sample in Galactic coordinates ($l$, $b$). The colour indicates distance, and the point size denotes $R_{\mathrm{PN}}$ ($10^{\prime\prime}$, $100^{\prime\prime}$, $500^{\prime\prime}$).}
\label{pn_all}
\end{figure}

\subsection{PNe spatial distribution} \label{subsec:spatial}

Contrary to the long-held view that the extinction–distance method could apply only within a few degrees of the Galactic plane \citep{2006Phillips,2016Frew,2022Kwitter}, we find that its limitation stems from data coverage. Previous PN extinction studies relied mainly on H$\alpha$ surveys (e.g. \hyperlink{G11}{G11}; \hyperlink{D21}{D21}), which cover only the Galactic plane (e.g. IPHAS, $|\mathit{b}|$<5$^\circ$; \citealt{2005IPHAS}). As shown in \autoref{pn_all}, Gaia’s all-sky coverage and improved method allow us to probe higher latitudes and even halo PNe. About 85.5\% of PNe lie within $|b|<10^\circ$, concentrated towards the plane. At higher latitudes, detections drop sharply, leaving larger, nearby PNe mostly within $\lesssim1$ kpc.

\subsection{PN radius--distance relation} \label{subsec:scale}

\begin{figure}[h!]
\centering
\includegraphics[width=0.9\hsize]{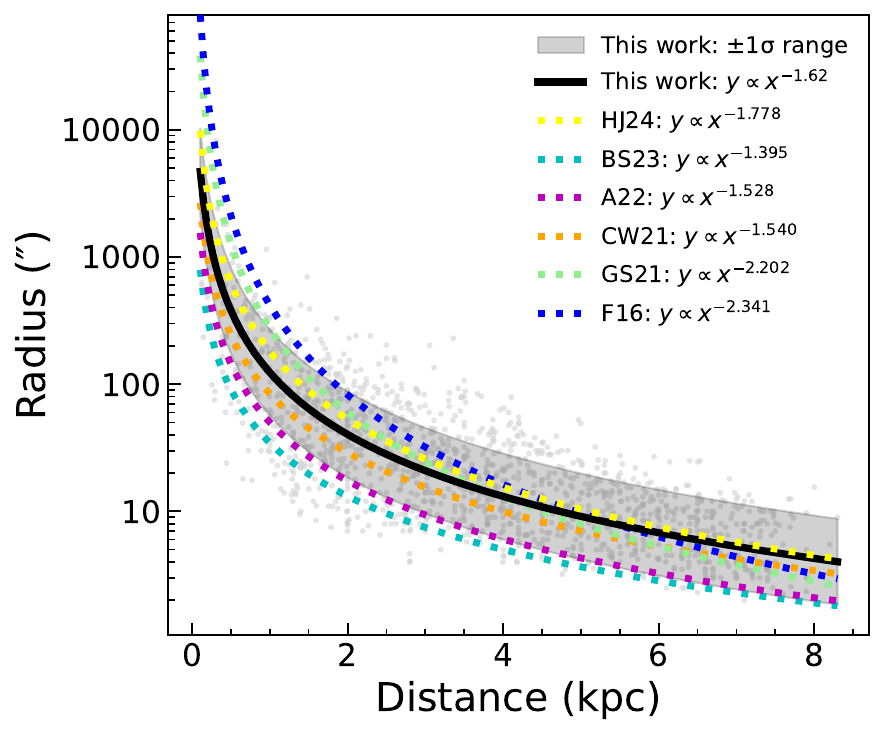}
\caption{Comparison of fitted $R_{\mathrm{PN}}$--distance relations. The grey points indicate our PNe. The black line and shaded area show our best fit and the 1$\sigma$ range. The dashed coloured lines represent literature fits obtained from their original complete datasets, using the same colour coding as in \autoref{combined_2x2}.}
\label{R_D}
\end{figure}

A clear trend between $R_\mathrm{PN}$ and distance is observed in our sample, with a power-law exponent of $\gamma=-1.62$. Applying the same procedure to the full datasets of \hyperlink{HJ24}{HJ24}, \hyperlink{BS23}{BS23}, \hyperlink{Ali22}{A22}, \hyperlink{CW21}{CW21}, \hyperlink{GS21}{GS21}, and \hyperlink{F16}{F16} produces exponents from $-2.3$ to $-1.3$ (mean residual 0.17), with our value at the median. \autoref{R_D} displays all $R_{\mathrm{PN}}$–distance relations. Our fit runs nearly parallel to the others, with all results lying within or near our $1\sigma$ range, confirming consistency with the established relation. In this work, the $R_\mathrm{PN}$–distance relation was not used as a distance scale, but only as a supplementary check in a very few exceptional cases. 

\begin{figure*}[h!]
\centering
\includegraphics[width=\hsize]{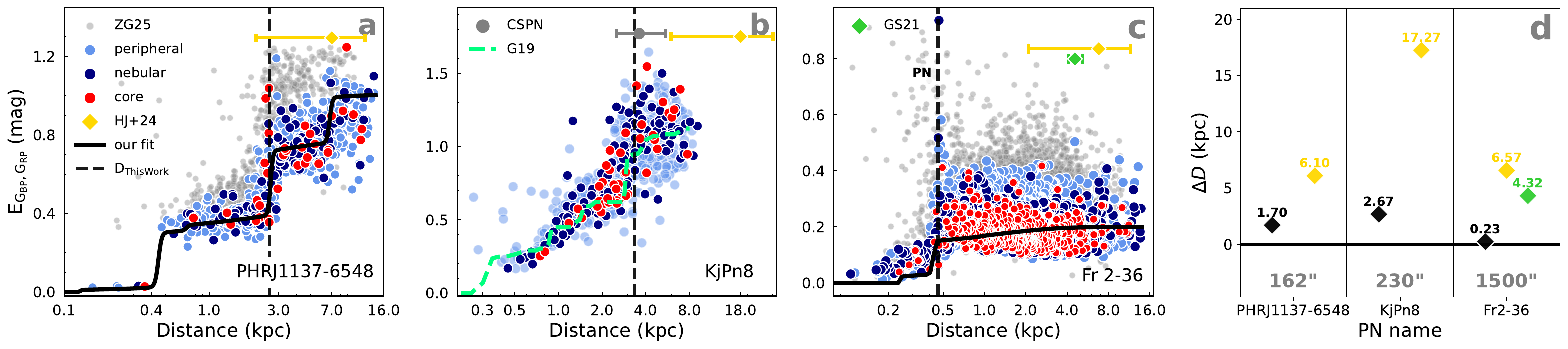}
\caption{Panels (a) - (c): Extinction–distance profiles for three PNe, with names labelled at the lower right. Sources and lines are colour-coded, as in \autoref{combined_2x2}. Panel (d): Residuals $\Delta D$ = $D$ - $D_{R_\mathrm{fit}}$, where $D_{R_\mathrm{fit}}$ is the distance predicted by the $R_{\mathrm{PN}}$–distance fit for each PN. PN names are listed along the x-axis, with the adopted literature value of $R_{\mathrm{PN}}$ indicated below each PN. The data points adopt the colours from panels (a) - (c), with our determinations highlighted as black diamonds. All residuals are labelled above their respective points.}
\label{pn_abcd}
\end{figure*}

\subsubsection{Applications: PN distance diagnostics} 
\label{subsubsec:r-d-2}
As noted in Section~\ref{subsec:Step3}, when distance constraints are insufficient to estimate the PN distance, or when derived distances differ markedly from literature values (e.g. the grey points in \autoref{compare_dis}), the $R_{\mathrm{PN}}$–distance relation serves as a supplementary diagnostic. The applications discussed below are illustrated by three representative cases (panels a–c of \autoref{pn_abcd}), with the corresponding $R_{\mathrm{PN}}$–distance residuals shown in panel d.

\hyperref[pn_abcd]{Figure~\ref{pn_abcd}a} shows two extinction jumps within the constraint range of \hyperlink{HJ24}{HJ24}, illustrating a complex sightline in which conventional methods fail to uniquely locate the PN. The $R_\mathrm{PN}$–distance relation favours the first jump as the plausible solution (residual 1.7 vs 6.1).
\hyperref[pn_abcd]{Figure~\ref{pn_abcd}b} presents a PN exhibiting a large methodological discrepancy, where our estimate differs from \hyperlink{HJ24}{HJ24} by a factor of 5.3. Observational limits prevent detection at the \hyperlink{HJ24}{HJ24} position. The $R_\mathrm{PN}$ check (panel d) favours our distance (residual 2.67) over \hyperlink{HJ24}{HJ24} (17.27; $\gg 3\sigma$ from the relation, with a 68\% distance uncertainty), demonstrating robustness for large-distance deviations.

\hyperref[pn_abcd]{Figure~\ref{pn_abcd}c} shows PN Fr 2-36 whose distance differs significantly from \hyperlink{GS21}{GS21} and \hyperlink{HJ24}{HJ24} (by factors of 10 and 13, respectively), likely due to a misidentified CSPN. First, the PN lies at high Galactic latitudes ($\sim23^\circ$) along a clean, low-extinction sightline free from background contamination. The single, sharp extinction jump at $\sim0.49$ kpc (maximum $\Delta E_\mathrm{G_{BP},G_{RP}}\approx1$ mag), consistent with the PN core and the trend in \hyperlink{ZG25}{ZG25}, thus reliably traces the PN. Second, only one of the four required priors (\hyperlink{GS21}{GS21}) applies to the \hyperlink{HJ24}{HJ24} distance estimate for this PN, yielding a 74\% uncertainty. The resulting overlap with \hyperlink{GS21}{GS21} therefore reflects a shared prior rather than independent evidence for distance reliability. Finally, our distance minimises the $R_\mathrm{PN}$–distance residual (0.23 kpc), while the literature values deviate by $\gg 3\sigma$ (6.57 and 4.32 kpc). With $R_\mathrm{PN}=1500\arcsec$ adopted from HASH, as also used in \hyperlink{CW21}{CW21}), a $\sim$5 kpc distance would imply an unphysically large PN. All indications strongly suggest that this CSPN has been misidentified.

\subsubsection{Applications: Discordant CSPN identifications} \label{subsubsec:r-d-3}
When \hyperlink{CW21}{CW21} and \hyperlink{GS21}{GS21} assigned different CSPNe to the same PN, we selected the most plausible one based on (1) stellar parameters ($T_{\rm eff}$, $\log g$) consistent with white dwarfs, (2) consistency among parallax distance, extinction jumps, and published constraints, and (3) proximity to the $R_{\mathrm{PN}}$–distance relation. A CSPN was regarded as well-supported if it met at least two of these criteria. Among 33 such PNe, we identified the more reliable CSPN in 15 cases (\autoref{15CSPN}).

\begin{figure}[h!]
\centering
\includegraphics[width=0.955\hsize]{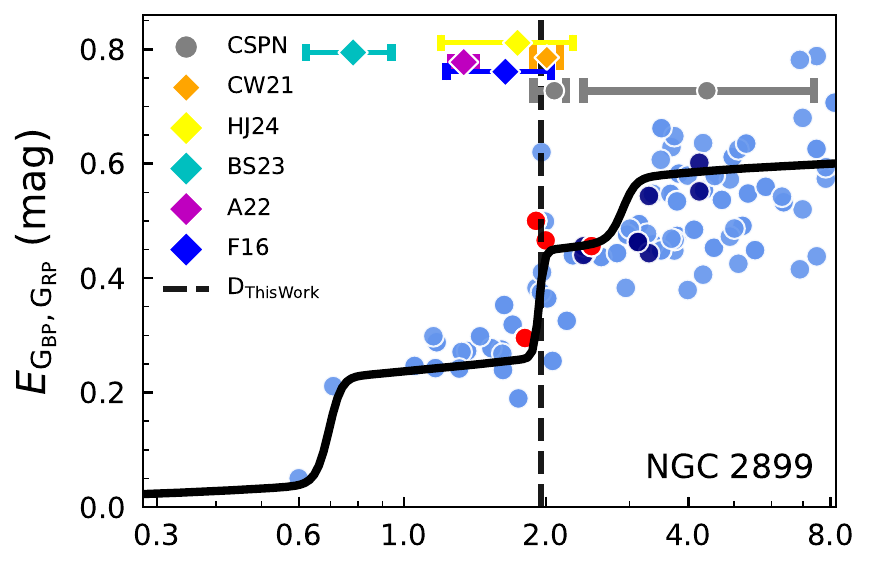}
\includegraphics[width=0.95\hsize]{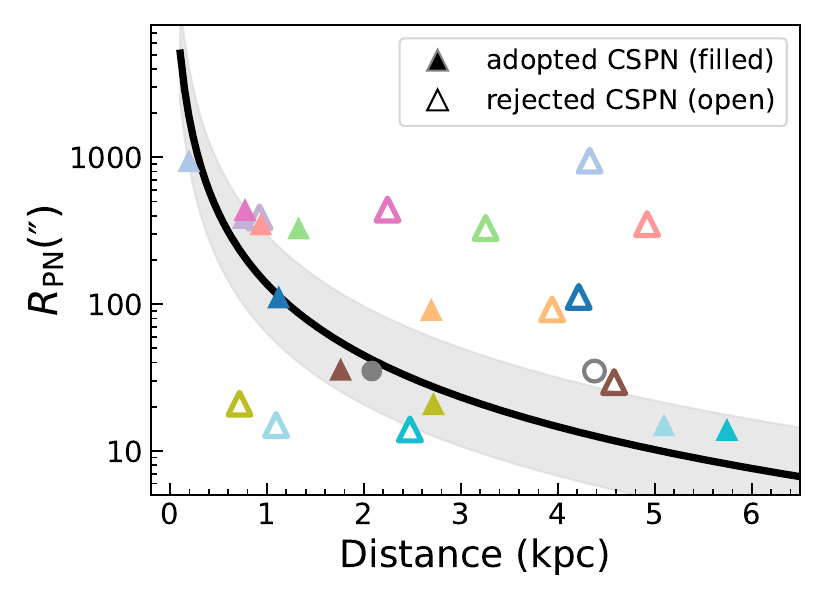}
\caption{CSPNe validation. Top: Extinction–distance profile of NGC 2899. The sources, reference distances, and lines follow the colour coding in \autoref{combined_2x2}. The two published CSPNe are shown in grey with error bars. Bottom: Discordant CSPNe. The black line represents the fitted $R_{\mathrm{PN}}$–distance relation with the shaded area marking the $\pm 1\sigma$ range. All PNe from \autoref{15CSPN} are included. The filled symbols indicate the adopted CSPN, and the open symbols indicate other published CSPNe. NGC 2899 is shown in grey circles, whereas CSPNe in other PNe are shown as coloured triangles.}
\label{2CSPN}
\end{figure}

\begin{table}[h]
\centering
\caption{Selected PN central stars with Gaia DR3 IDs}
\label{15CSPN}
\begin{tabular}{lc}
 \hline \hline
PN name & Gaia Source ID  \\
\hline
NGC 6445 & 4119025030278036608 \\
FP J1824-0319 & 4269678120544342144 \\
Ou 3 & 2020279846433107200 \\
Pa 4 & 2233941236591666688 \\
Sh 2-176 & 421836329014389248 \\
Sh 2-188 & 509206447837376128 \\
EGB 6 & 615161091995252864 \\
NGC 2440 & 5716990599321529216 \\
FP J0905-3033 & 5635240088724202496 \\
Fr2-6 & 5314960460454288768 \\
NGC 2899 & 5307241888842877696 \\
PHR J1123-6030 & 5337515429678393984 \\
PHR J1309-6457 & 5858960227979564800 \\
Th 2-A & 5865192362963086592 \\
PHR J1721-5122 & 5925231465889406720 \\
\hline
\end{tabular}
\end{table}

As an illustration, in the top panel of \autoref{2CSPN}, the two CSPNe of NGC 2899 can be distinguished using the latter two criteria. An analysis of extinction jumps shows three jumps along the line of sight. One CSPN lies at 2.0 kpc with a 5.9\% uncertainty, precisely matching the second jump. This jump encloses the core star, rises sharply by 0.4 mag, and is consistently covered by three independent distance measurements, showing good agreement. 
The other CSPN is at 4.4 kpc with a 69\% uncertainty, with its error range covering only the third jump. The 2.0 kpc CSPN also has a smaller residual with respect to the $R_{\mathrm{PN}}$–distance relation, indicating that it is the more reliable CSPN. The bottom panel shows the $R_{\mathrm{PN}}$–distance relations for all PNe listed in \autoref{15CSPN}, shown as triangles, with the PN discussed here indicated by circles. Each PN is shown in the same colour for its associated CSPNe. In all cases, the adopted CSPN (solid dot) lies closer to the fitted relation than the other (open circle), confirming the internal consistency of our identifications and the robustness of the $R_{\mathrm{PN}}$–distance relation.

Collectively, these tests confirm the reliability and internal consistency of the method. They demonstrate that the extinction–distance approach provides a robust basis for constructing a large and homogeneous PN distance catalogue.


\section{Summary and conclusions}\label{Conclusions}
We presented an optimized extinction-distance method based on Gaia data, providing a CSPNe parallax-independent and physically motivated framework for determining distances to PNe with confirmed CSPNe. Our main results are as follows:

\begin{enumerate}
    \item We provided extinction-based distance estimates for 1,066 PNe, including 124 objects measured for the first time. The relative distance uncertainty has a median value of 13\% and is below 20\% for 87\% of the sample. Among them, 765 CSPNe have relative parallax errors exceeding 20\%, and 128 lack parallax measurements, demonstrating that our method provides a robust complement when parallaxes are uncertain or unavailable.

    \item We significantly extended the reliability and applicability of traditional extinction–distance measurements, enabling distance determinations for high-Galactic-latitude PNe. The sample of PNe with individually measured distances increases from a few dozen to over one thousand, forming the largest catalogue of extinction-based PN distances to date.
    
    \item We examined 33 PNe with multiple and inconsistent CSPNe identifications, recommend more reliable CSPNe for 15 objects, and find that the uniquely reported CSPN of Fr2-36 is likely misidentified.
\end{enumerate}

Future work will further refine extinction estimates, expand the sample, and incorporate forthcoming Gaia DR4 data to investigate the extinction properties and local environments of PNe in greater detail.

\section*{Data availability}

The resulting homogenized PN distance catalogue is available at the CDS via the VizieR online service and on Zenodo at \url{https://zenodo.org/records/17010405}.
\begin{acknowledgements}
We sincerely thank the anonymous referee for the careful reading of the manuscript and for the very constructive and helpful comments, which we greatly appreciate. We thank Prof. Xiaodian Chen for very helpful discussions and suggestions. 
This work is supported by the National Natural Science Foundation of China (NSFC) projects 12133002 and 12373028. S.W. acknowledges the support from the Youth Innovation Promotion Association of the CAS (grant No. 2023065). This work has made use of data from Gaia survey.
\end{acknowledgements}

%

\bibliographystyle{aa}
\bibliography{reference}


\end{document}